# FIRST POINT-SPREAD FUNCTION AND X-RAY PHASE-CONTRAST IMAGING RESULTS WITH AN 88-mm DIAMETER SINGLE CRYSTAL

A. H. Lumpkin[1#], A.B. Garson[2], and M.A. Anastasio[2]

[1]Fermi National Accelerator Laboratory*, Batavia, IL 60510 USA, [2] Department of Biomedical Engineering, Washington University in St. Louis**, St. Louis, MO 63130 USA

## ABSTRACT

In this study, we report initial demonstrations of the use of single crystals in indirect x-ray imaging with a benchtop implementation of propagation-based (PB) x-ray phase-contrast imaging. Based on single Gaussian peak fits to the x-ray images, we observed a four times smaller system point-spread function (PSF) with the 50-$\mu$m thick single crystal scintillators than with the reference polycrystalline phosphor/scintillator. Fiber-optic plate depth-of-focus and Al reflective-coating aspects are also elucidated. Guided by the results from the 25-mm diameter crystal samples, we report additionally the first results with a unique 88-mm diameter single crystal bonded to a fiber optic plate and coupled to the large format CCD. Both PSF and x-ray phase-contrast imaging data are quantified and presented.

Index:

*Work at Fermilab supported in part by Fermi Research Alliance, LLC under Contract No. DE-AC02-07CH11359 with the United States Department of Energy.

**Work at Washington University in St. Louis was supported in part by NSF award CBET1263988 and NIH awards EB023045 and EB020604.

#lumpkin@fnal.gov

# I. INTRODUCTION

X-ray phase-contrast (XPC) imaging is an emerging technology that holds great promise for biomedical applications due to its ability to provide information about soft tissue structure [1]. A variety of XPC imaging methods continue to be actively developed and investigated for characterizing soft tissue or biological samples that present limited x-ray absorption contrast. Such methods include propagation-based (PB) imaging [2], crystal analyzer-based imaging [3,4], grating-based imaging based on the Talbot or Talbot-Lau effect [5], and edge-illumination imaging [6]. A grating-based technique has recently been employed in a pre-clinical mammography study with 21-keV x-rays [7]. Of the available techniques, PB XPC methods are the simplest to implement since they don't require optical elements between object and detector. For this reason, PB XPC methods are also intrinsically more dose efficient [8,9]. In addition, PB XPC can be implemented with a much lower degree of x-ray beam partial coherence [10]. Recently, the availability of high-brilliance micro-focus sources, which are based on a liquid-metal anode instead of a solid target [11,12], have enabled new applications of PB XPC imaging in the laboratory [13]. A technical requirement of PB XPC imaging is that a high spatial resolution x-ray detector is employed, which can directly record the XPC-induced variations in the wavefield intensity. There remains a need to develop improved detector technologies that can provide high spatial resolution while maintaining detection efficiency and field-of-view.

To address this need, based on results on imaging of relativistic electron beams with single crystals [14], we proposed transferring single-crystal imaging technology for use with XPC imaging. We report initial indirect x-ray imaging tests that demonstrated improved spatial resolution with single crystals compared to the $Gd_2O_2S$:Tb polycrystalline phosphor (P43) in a commercial, large-format CCD system. Using a microfocus x-ray tube as a source of 17-keV x-rays and the exchangeable phosphor feature of the camera system, we compared the point-spread function (PSF) of the system with the reference phosphor to that with several rare-earth-garnet single crystals of varying thickness and with 25-mm diameters.

In x-ray imaging applications, a tradeoff exists between scintillator screen spatial resolution and detector efficiency. This has been particularly true for polycrystalline phosphors such as $Gd_2O_2S$:Tb where the resolution at full-width-at-half-maximum intensity (FWHM) for 10- to 20-keV x-rays is approximately equal to the screen thickness [15]. Generally, a thinner screen is employed when improved resolution is needed with the concomitant decrease in detector efficiency. If an Al reflective coating is added on the front surface for light collection efficiency, this will also impact the resolution. It has been observed [14,16] that superior resolution can be obtained with single crystal scintillators of comparable thickness, although there still may be a trade on efficiency depending on the materials used. In this work, the effects of the reflective coatings and fiber-optic plate (FOP) depth-of-focus (DOF) terms on spatial resolution obtained with such phosphors are also investigated. Subsequently, a custom-ordered 88-mm diameter YAG:Ce single crystal bonded to a FOP was installed in a large format CCD system. The first indirect x-ray imaging tests to evaluate the system PSF and the first examples of PB XPC imaging with this improved system resolution have been performed and are presented.

## II. EXPERIMENTAL TECHNIQUES

In this section, we describe the path from the electron-beam imaging application to the x-ray imaging application using single crystals, the features of the single crystals, source, and methods.

### A. Electron-beam Imaging Background

The improved spatial resolution with single crystals over polycrystalline or powder samples had been previously noted in the imaging of relativistic electron beams [14]. Examples are shown in Fig. 1 as deduced from results at various accelerator laboratories [17-22], and the concept is being applied to indirect x-ray imaging in this research. The powder or polycrystalline data points were extracted from previous reports using a direct comparison to alternative beam profiling techniques based on optical transition radiation imaging or wire scanners in the tests. We note Chromox is a trade name for chromium-doped aluminum oxide which has been implemented in the past years more for its robustness to beam irradiation than its spatial resolution. In all cases the derived resolution sigma in microns for powder data fall at noticeably larger values than the single crystal data available at that time. It should be noted that even with a grain size of 5 μm in the polycrystalline YAG:Ce and YAG:Tb examples, the spatial resolutions are still 50-80 μm due to the light scattering off the multiple grain surfaces within the sample. This particular light-scattering effect persists in x-ray imaging as it is inherent to such phosphor samples, while it is avoided in single-crystal scintillators.

The rare-earth-garnet single crystals of 25-mm diameter were both procured from a vendor and borrowed from the Fermi National Accelerator Laboratory (FNAL) and Argonne National Laboratory (ANL) linear accelerator labs. For these studies, we have chosen a set of single crystals of varying thickness, with and without the presence of a fiber optic plate (FOP). The crystal types were cerium-doped yttrium aluminum garnet, YAG:Ce, and lutecium aluminum garnet, LuAG:Ce. We obtained thicknesses of 50, 100, and 200 μm. Two paired samples were also obtained from Crytur with Al coating as an optical reflector for the 50- and 100-μm examples. These were used to assess the role of the input FOP's depth of focus on the system PSF and signal output.

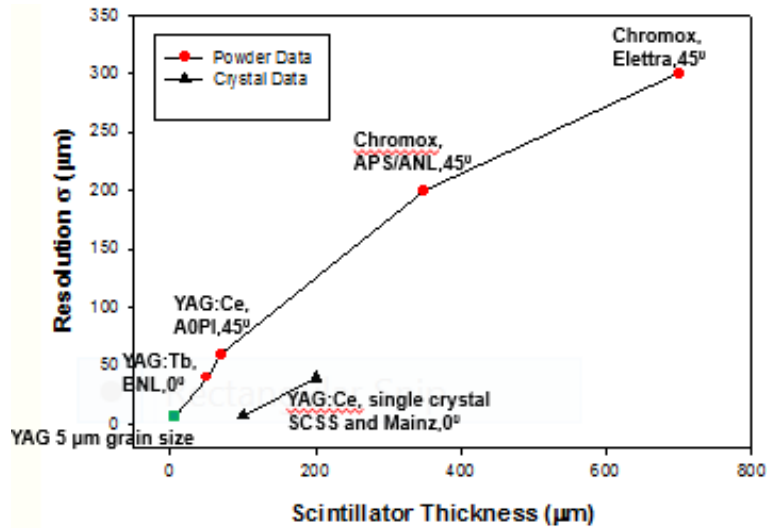

Figure 1: Combined plot of powder and crystal screen thicknesses and deduced spatial resolution terms based on a quadrature analysis of the reported observed image sizes with the OTR or reference image size [14]. The data points with relativistic electron beams are collected here for the following samples: 1) $Al_2O_3$:Cr (Chromox) at Elettra [17], 2) $Al_2O_3$:Cr at ANL [18], 3) YAG:Ce at Fermilab [14], 4) YAG:Tb at BNL [19], and 5) our estimate for a screen at the 5-μm grain size at 45 degrees [20]. The scintillator crystal points are from 6) 100-um thick YAG:Ce at SCSS [21] and 7) 200 um thick YAG:Ce at Mainz [22], in both latter cases the crystal surface plane was at 90 degrees to the beam direction while samples 1, 2, and 3 were at 45 degrees. The lines between points are used to guide the eye.

One of the final objectives was to obtain first indirect x-ray images from the 88-mm diameter crystal bonded to the 90-mm diameter FOP. This unique crystal was also prepared by Crytur,Inc. under a custom order from FNAL, and it is shown in comparison to a quarter in Fig. 2a. Another perspective is shown in Fig. 2b with the hand-held sample. We will report the first point-spread function (PSF) tests of this sample in a fiber-optic coupled configuration while a lens-coupled configuration is reported elsewhere using the x-ray beam in a synchrotron radiation source beamline [23]. This scintillator crystal's combined high resolution and large diameter make it a candidate for potential use in PB x-ray phase contrast imaging with a large format CCD system [24], x-ray crystal diffraction studies, or wafer topography studies.

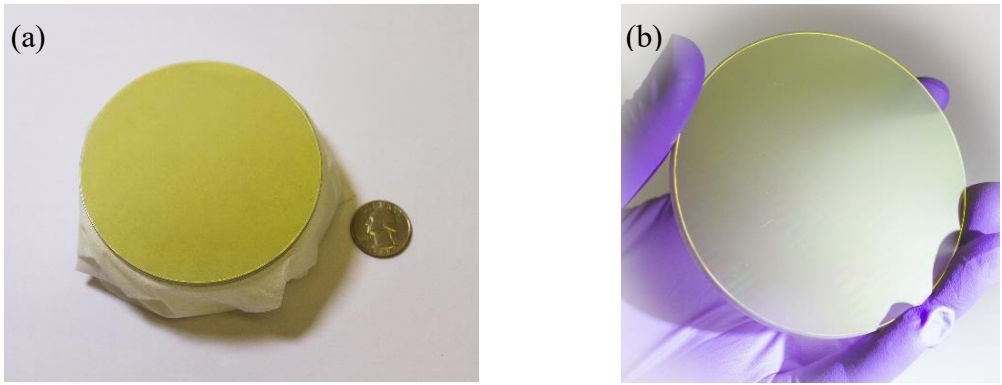

Figure 2: Comparison of the 88-mm diameter crystal to a quarter (a) The latter crystal is more than 12 times bigger in area than a 25-mm diam standard crystal. (b) Photograph of the 88-mm diameter crystal (a.k.a. "Katherine's Krystal") on the FOP (Photos by E. McCrory, FNAL)

*B. XPC Imaging Setup*

X-ray imaging experiments were conducted at Washington University in St. Louis [25]. The imaging setup is comprised of a microfocus x-ray source, a high precision stage and rail system, and a high-resolution Princeton Instruments (PI) Quad-RO x-ray camera as schematically shown in Fig. 3. The x-ray source is a Kevex PXS10-65W with cone beam, tungsten anode, 7-100 micron spot sizes, and 45-130 kV tube voltages. The Quad-RO-4096 camera is a Peltier-cooled (-40 degrees C) CCD, with 15-micron pixel pitch for a 4096 x 4096 array [26]. It has 14-bit intensity quantization and a PSF to be determined (generally 33-40 microns was ascribed).

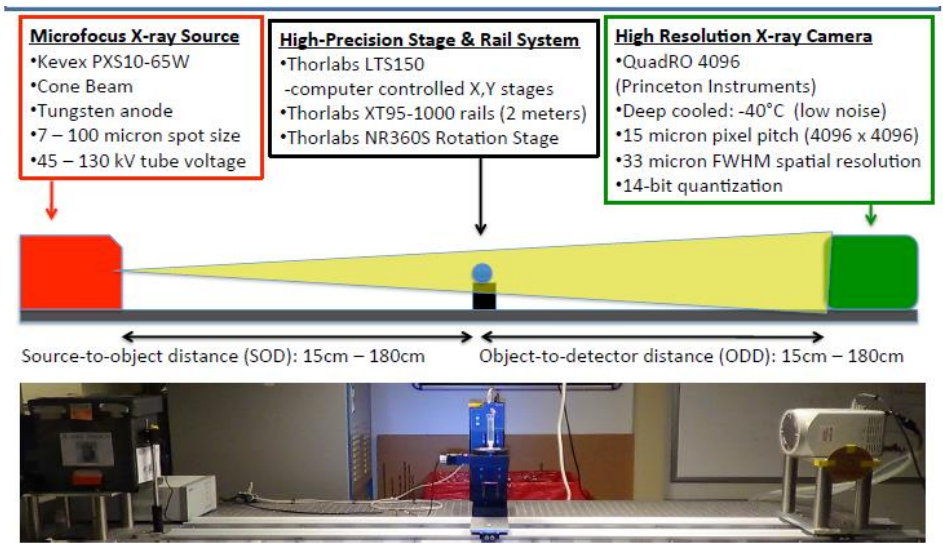

Figure 3: Schematic and photograph of the Washington University XPC imaging laboratory showing the microfocus x-ray source, the sample translation stages, and the Quad-RO-4096 camera [25].

*C.     Methods*

In order to evaluate the system PSF, we used collimated beams whose smallest spatial extent were 2-3 times smaller than the expected PSF value (since a resolution pattern made by sandwiching thin foils of known thickness was not practical at this scale as done in the past [27]). We placed sequentially the collimators from an Amptek set on a lead plate with a hole drilled in it smaller than the W disk diameter. This plate was leveled with shims against the outer flange surface of the Quad-RO camera and positioned for the x-ray images to fall in the central area of one of the four quadrants of the CCD array. The set included collimators of 400, 200, 100, 50, and 25 μm in diameter. We estimated the rms size from a simple calculation of the effective root-mean-square (rms) size of an aperture of width w. This is the SQRT of the integral of $x^2 dx$ with limits from -w/2 to w/2 normalized by 1/w. This gives w/SQRT 12 for the rms value, and then we multiplied by 2.35 to obtain the FWHM of an assumed Gaussian profile for the effective x-ray source size as shown in Table I.

Table I.  Summary of the x-ray collimators' features used to assess the PSF with 17-keV x-rays.

| Collimator # | Diameter (μm) | FWHM (μm) |
|---|---|---|
| 1 | 400 | 272 |
| 2 | 200 | 136 |
| 3 | 100 | 68 |
| 4 | 50 | 34 |
| 5 | 25 | 17 |

The x-ray source was operated at 7-μm spot size, at 25- or 60-kV tube voltage, at 150-μA current, 3.8 W, and located 0.9 m from the camera. Typical image integration times were 30s, and we averaged over five frames.  Data were dark current subtracted, but only any mesh images were flat-field corrected. They were acquired and displayed with PI software [28]. Typically, only a 400 x 400 channel region of interest (ROI) was then selected for processing in FNAL's image processing program, ImageTool, a MATLAB based program [29]. This program fits Gaussian profiles to the projected profiles from the selected smaller ROI and provides the amplitude, mean position, sigma, and the corresponding errors for each from the analyses. Background fit options are linear, flat, and quadratic. The program can fit up to 8 different peaks in the ROI, and we used this feature to assess the modulation of the wire grid and mesh data (not shown).  It also provided the option for fitting projected profiles to a double Gaussian when that issue arose.

### III. X-RAY IMAGING RESULTS

The x-ray imaging results include the evaluations of the system PSF with collimated x rays, the fiberoptic depth-of-focus effect, the crystal efficiency results, the light-scattering term, and the first PB XPC images with the 88-mm diameter single crystal.

#### A.  Collimated X-ray Image results

We show the initial results of the 50-μm diameter collimator images as an example in Fig. 4. We obtained the reference Al–coated P43 phosphor data first and immediately noted that the projected vertical profile of 88 μm (FWHM) in Fig. 4a indicated the PSF was larger than the expected 40 μm. After taking the whole collimator set data with the P43, we installed the YAG:Ce and LuAG:Ce 50-μm thick crystals in the QuadRO positioned over two diagonal quadrants of the 4-quadrant sensor. Both the YAG:Ce and LuAG:Ce crystals had image sizes of about 36±1 μm (FWHM) as in Fig. 4b, very close to the calculated aperture FWHM for this collimator. Figure 5 shows a summary plot comparing the polycrystalline and single crystal results. Using the smallest aperture of 17-μm FWHM (case 5), we deduced the system PSF (found by subtracting out the aperture size in quadrature) was about 21 μm with the single crystals, *4 times smaller* than that with the reference P43 phosphor.

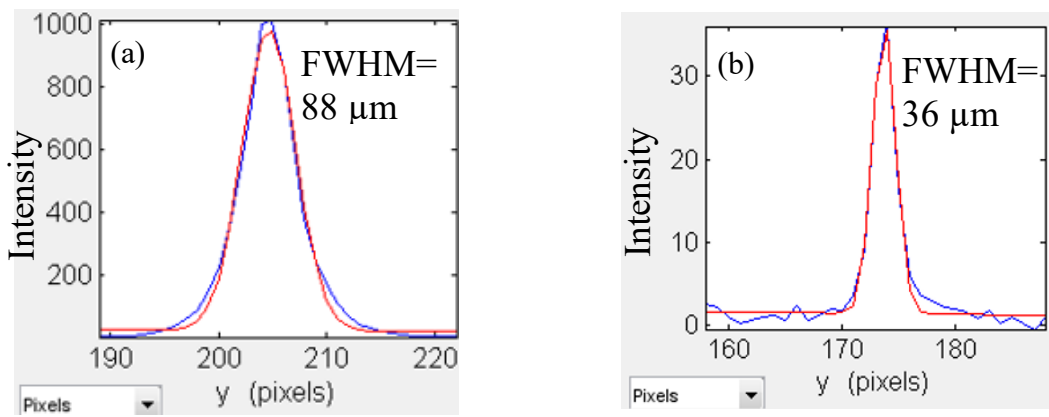

Figure 4: Initial images using the 50-μm diameter collimator with a) the reference P43 phosphor and b) a 50-μm thick single YAG:Ce crystal.  Projected profile data (blue) and fitted Gaussian curves (red) are compared.

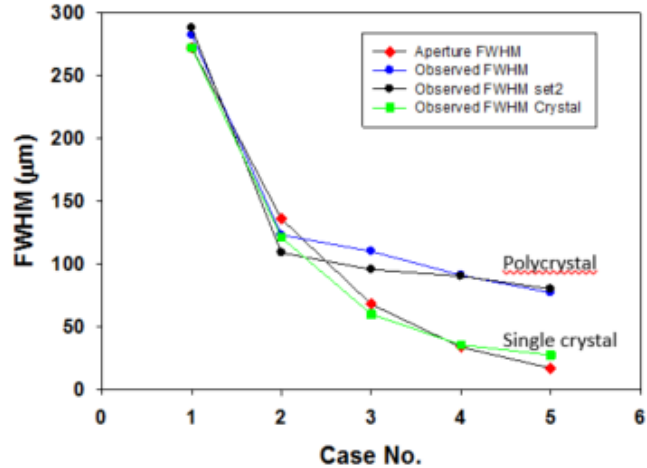

Figure 5: Plots of the measured projections for the different collimated images. The single crystal data clearly show better spatial resolution than the polycrystal with the smallest collimators.

In a subsequent test series, we obtained two Al-coated samples, one on a 50-µm thick and one on a 100-µm thick YAG:Ce crystal. In Fig. 6 we show that the depth of focus (DOF) of the input FOP plays a role in imaging with increasing effective optical thicknesses (which accounts for the increased path length by a factor of 2 with the Al reflector involved). The system PSF grows from 21 µm with 50-µm crystal thickness to about 70 µm with an effective crystal thickness of 200 µm. So scintillator efficiency and this DOF term still need to be considered as a trade. The DOF term is dominating the system PSF at 100-200 µm thicknesses for a single crystal while it is comparable to the light scattering term of the P43 with 68-µm effective thickness. We note the bonded 100-µm-thick crystal had a PSF of about 10 µm in a lens-coupled x-ray imaging test [23].

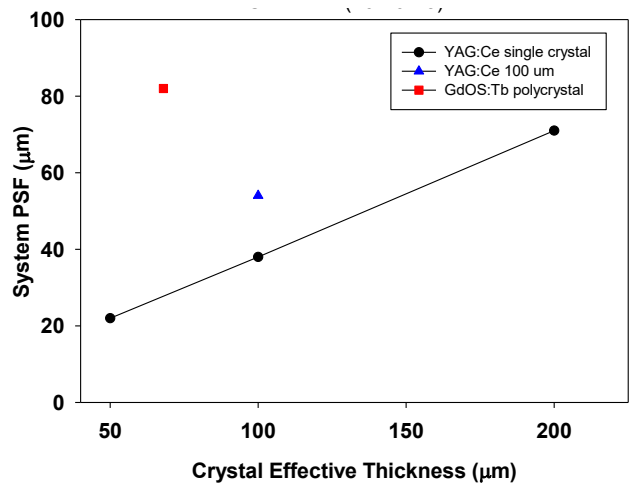

Figure 6: Plot of the system PSF vs. the effective crystal optical thickness showing the FOP depth-of-focus effect.

### B. Collimated X-ray Image results: Reflective Coatings

In addition, we have also revisited the collimated images for investigation of the effects of the Al reflective coating on the front surface which is typically added to boost light-collection efficiency. The scintillator radiates into all angles, and the backward directed light is redirected forward by the Al. Although, it is acknowledged that there is some loss in spatial resolution, we suggest our data show the magnitude of this trade. For a given polycrystalline phosphor thickness, the addition of the reflected light component doubles the distance over which light scattering occurs.

The geometry of the system is schematically shown in Fig. 7. If the coupling is sufficient, then another term to consider for system PSF blurring is the phosphor with the Al reflector. As noted earlier, the single Gaussian fit profile missed matching the P43 experimental profile, being both lower at the peak and lower at the base. This is symptomatic of a narrower Gaussian peak sitting on a broader peak. As a hypothesis, the narrow peak is the forward light going to the FOP through the 34-μm thickness, and the broader peak would be the backward directed light generated all through the thickness which was then redirected/reflected forward to the CCD. This light effectively encountered more of the polycrystalline scattering as well as originating from "roundtrip" distances of 35-68 μm. These larger source distances would be even further outside of the limited depth of focus of the FOP and contributed to additional blurring of the image. The signal is increased as expected, but the resolution is degraded in this case by about a factor of two.

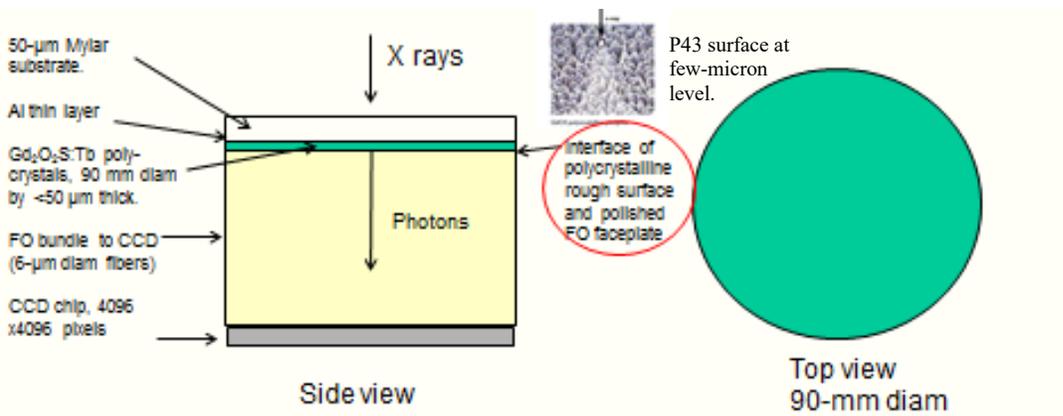

Figure 7: Schematic of the camera arrangement showing the mylar substrate, Al layer, scintillator, FO bundle, and CCD chip.

We show in Fig. 8 an example of the Run 2 polycrystalline data with 25-μm collimator aperture. At the left, is the single Gaussian fit result with sigma of 2.29 pixels corresponding to a y-profile size of 81 µm (FWHM). At the right, the two-Gaussian peak fit results in a much better match to the data at the peak and the base. The two Gaussians are centered to less than 0.1 pixels of each other with similar peak amplitudes, but the narrow one is 42.3 µm and the broad one is 117 µm FWHM. The narrow one is very close to the PI web site value taken with a 5-µm diameter pinhole [9]. If one subtracts the aperture size in quadrature from the narrow peak size we obtain a PSF of 39 µm, very close to the 40-µm PI value for a 17-keV optimized P43 phosphor.

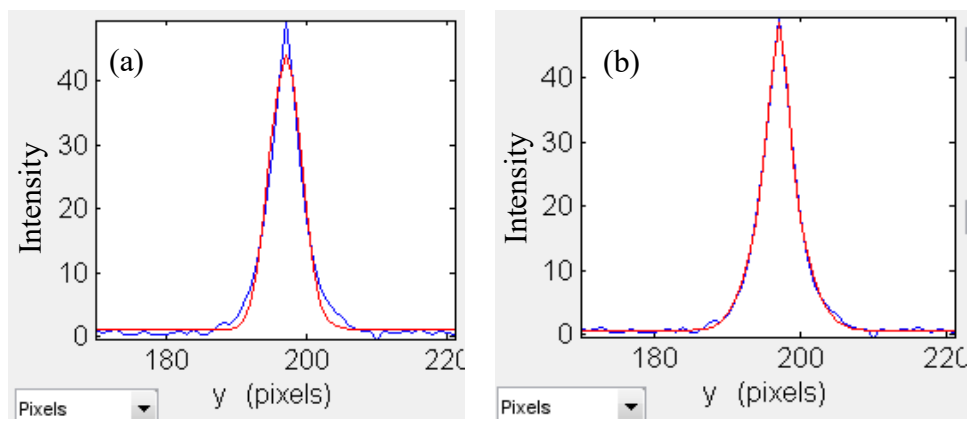

Figure 8: Comparison of the 25-µm diameter collimator image processing using the polycrystalline data (blue) from the Run 2 recoupling. (a) The image and single Gaussian peak fit (red) to the vertical profile and (b) the same profile with two-Gaussian peak fit are shown.

We performed similar procedures for the 50- and 100-µm aperture data. The three narrower Gaussian fit points (burgundy squares) from the two-peak fits are plotted in Fig. 9 as well as the single Gaussian fit curves for Run 1 (blue), Run 2 (green), and the Run 3 (black) single crystal results. The decreasing image size of the narrow peak with the aperture diameter decrease is encouraging, and the broad peak does also decrease with decreasing apertures. We also note the Run 3 values at 100 µm and 50 µm diameter apertures are lower than the values on Run 1 which would be consistent with improved coupling on Run 2. We postulate the FOP DOF term is also in play, effectively at 50-60 µm for the averaged reflected light. The polycrystalline light scattering term is at the 80-µm level for this P43 sample with Al reflector.

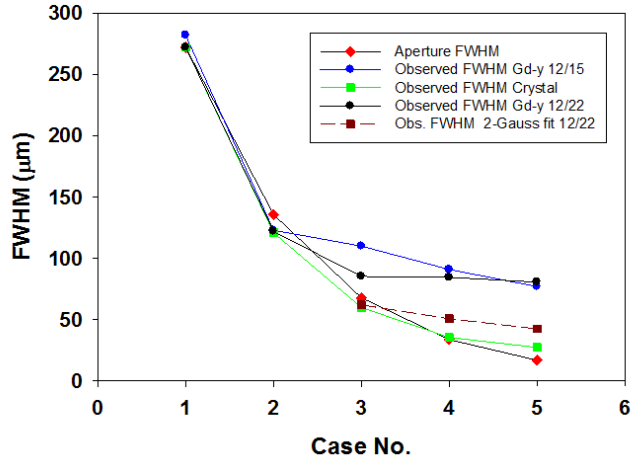

Figure 9: Comparison of the calculated aperture FWHM and the observed pinhole image vertical profile FWHM on Run 1 (from Fig. 5), Run 2 (single crystals), Run 3 single Gaussian, and Run 3 narrow Gaussian fit from 2-Gaussian peak fit.

The system PSF seems to limit at about 80 μm with the polycrystalline phosphor when using single Gaussian fits to the image's projected profiles on the y axis. This is basically the average of the 42-μm term and the 120-μm reflected term per the two-Gaussian term hypothesis. However, the hypothesized narrow peak fit limits at 42 μm with the smallest aperture while the Quad-RO system with the single crystals installed has an estimated PSF of 21 μm.

## B. Crystal Efficiency Results

The aspect of crystal efficiency vs. thickness for 17-keV x rays was evaluated using the product of the peak height and sigma values from the Gaussian-peak fits to the collimated aperture data for the different crystal samples. The data shown in Fig. 10 indicate the signal approximately doubles linearly for crystal thicknesses from 50 to 100 μm for the three, fixed-aperture sizes, but the signal gain is only ~50% as one increases the effective crystal thickness from 100 μm to 200 μm. All samples were using the YAG:Ce scintillator. These are relative measurements, but the LuAG:Ce crystals for the same thickness were found to be 2-3 times more efficient than the YAG:Ce. The YAG:Ce efficiency is about 8-12 % of the P43 reference phosphor in this case so the LuAG:Ce would be preferred in general.

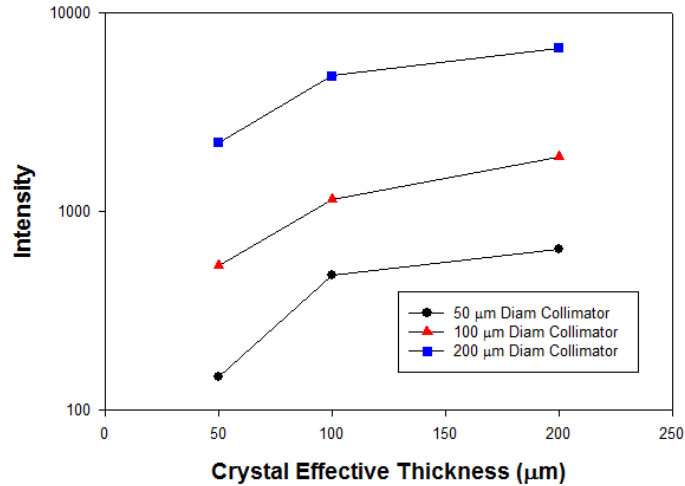

Figure 10: Plot of signal intensity vs. YAG:Ce crystal thickness using the image projected profile parameters for three collimator sizes. At these x-ray energies, the signal increase benefit was reduced when using larger than 100-μm thicknesses.

### C.  88-mm Crystal PB-XPC Imaging Tests

We procured an 88-mm diameter YAG:Ce crystal bonded to a 90-mm FOP as shown in Fig. 2 [30].  Based on our initial single-crystal tests, we had opted for a 100-μm thick YAG:Ce crystal since the more efficient LuAG:Ce was not yet available at such a large diameter. This increase from a 50-μm thickness was done to increase the efficiency with acceptable loss in spatial resolution. We left the Al reflective coating as an option to consider after first tests. This large crystal/FOP was installed in the QuadRO-4096 camera for evaluation with a suitable phantom of bioimaging relevance. For the PB-XPC tests, the source to object distance was 24 cm, the object to detector distance was 40 cm, and 30s exposures were used with the Kevex source. At the time of these tests, the CCD had 2x2 binning for a 30-μm pixel pitch. This step actually limited the resolution of the final system PSF with the single crystal so in future tests we plan for 1x1 binning again. As a point of reference, we first used XPC imaging on a randomly dispersed set of 33-μm diameter carbon fibers. These are clearly displayed in the zoomed-in image in Figure 11. Note, with the geometrical magnification of 2.66 in the PB-XPC geometry and 2x2 chip binning, the effective calibration factor was 12 μm/pixel. The observed fiber image sizes are 3-4 pixels or about 36-48 μm. This would include the system PSF contribution of about 30 μm added in quadrature with the fiber diameter.

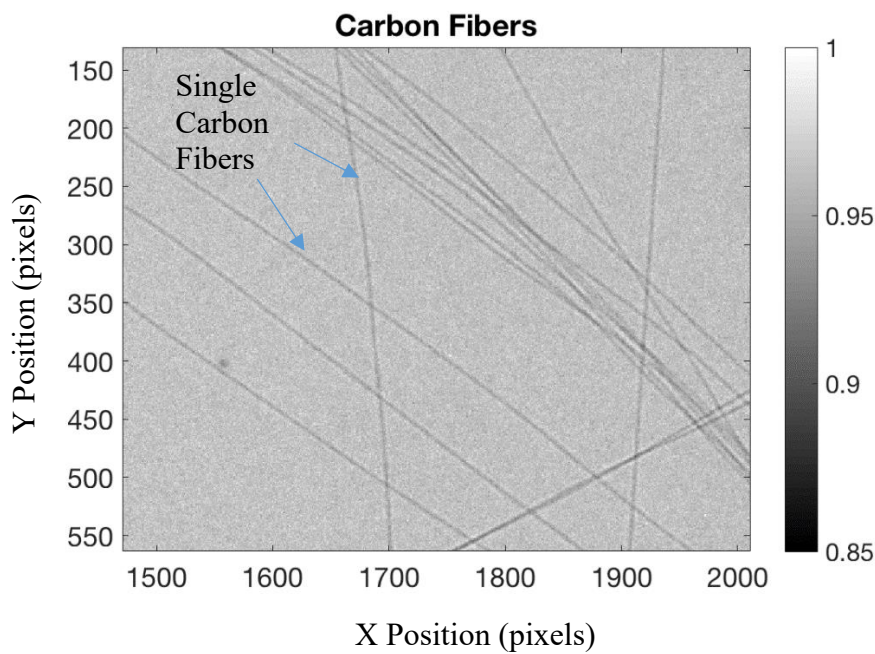

Figure 11: PB-XPC image of the 33-µm diameter carbon fibers randomly held in a plastic envelope. The intensity grey scale is shown in the bar at the right.

The second test object was a preserved dragonfly from the CBL inventory. Views in Fig. 12a,b are shown of a portion of the object's head, thorax, abdomen, wings, wing sockets to illustrate the difference in sharpness of the absorptive x-ray imaging (12a) and PB-XPC imaging (12b), using the large crystal for the first time. The test object was not identically oriented for the two configurations, but we endeavored for them to be close. The regions of interest (ROIs) are indicated in red, and the fine structure of the wings of the inset is seen more clearly with PB-XPC than absorptive in Fig. 12 d,c, respectively. These are partly a testament to the use of XPC imaging to identify finer details since the feature size and the 2x2 CCD binning may not fully display the crystal's advantage of ~30-µm PSF. The zoomed in images in Fig. 13 for absorptive and XPC data visibly show features more sharply in the XPC image of 13b than the absorptive image of Fig. 13a.

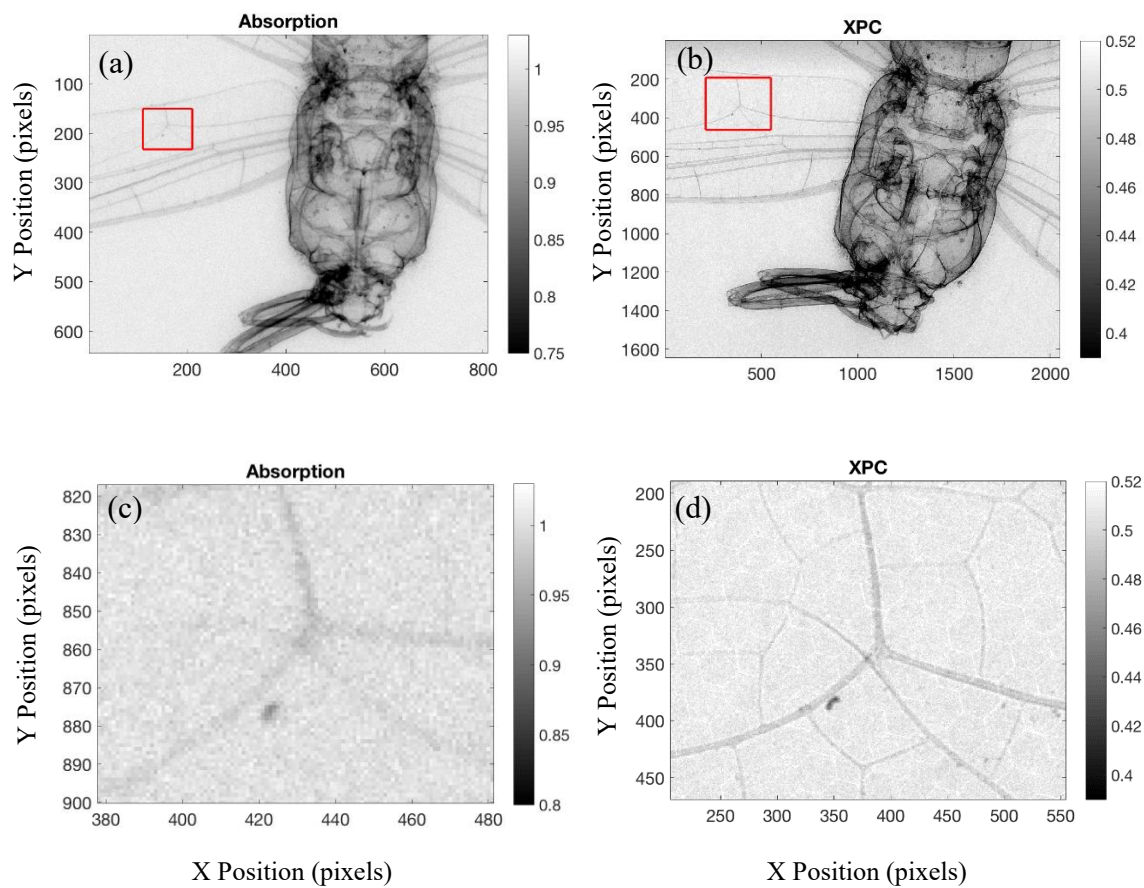

Figure 12: Comparison of the (a) x-ray absorptive and (b) PB-XPC images of a portion of a dragonfly subject obtained with the 88-mm diameter crystal installed in the indirect x-ray imaging system. The 4x4 mm$^2$ ROIs in red of each image are zoomed in and shown in (c) and (d), respectively. The intensity grey scale is shown in the vertical bar at the right.

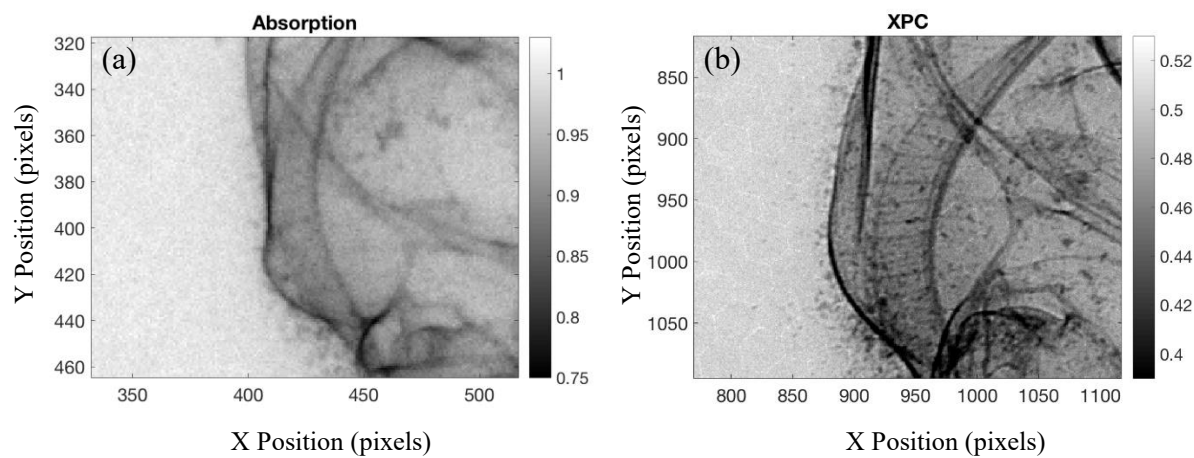

Figure 13: Comparison of the (a) x-ray absorptive and (b) PB-XPC images of a portion of a dragonfly subject obtained with the 88-mm diameter crystal installed in the indirect imaging system. Note the visibly sharper features in the XPC image at the right. The intensity grey scale is shown in the vertical bar at the right.

## IV. SUMMARY

In summary, we have performed initial studies of the improved spatial resolution obtained with rare-earth-garnet single crystals compared to the reference P43 sample in a commercial x-ray camera. We observed a 4 times better system PSF with the 50-µm thick YAG:Ce and LuAG:Ce single crystals. In the course of these investigations, we have elucidated the interplay of the FOP DOF term and the light-scattering term in the polycrystalline P43 sample. The latter indicates a spatial resolution limit similar to the effective thickness. The single crystal effective thicknesses were used to probe the depth-of-focus effect more clearly, and this study indicated it plays an increasing role starting at 50-100 µm in effective thickness. We hypothesized and quantified the deleterious effect on resolution of the Al coating at the phosphor. Finally, we obtained an 88-mm diameter single crystal bonded to a FOP and integrated this into a large-format camera for laboratory-scale PB-XPC imaging for the first time [30]. This option with 2x2 binning had about two times better PSF than the P43 reference system, but the former would be improved with the 1x1 binning option. This scintillator crystal's combined high resolution and large diameter make it a candidate for potential use in PB XPC imaging with a large format CCD system [24], x-ray crystal diffraction studies, or wafer topography studies. Further imaging tests of the installed large crystal with computed tomography techniques are also planned in the future.

## V. ACKNOWLEDGEMENTS

The first author acknowledges the support of E. Ramberg of PPD/FNAL and R. Dixon and N. Eddy of AD/FNAL and the loan of single crystals from D. Crawford of FAST/FNAL, W. Berg of APS/ANL, and M. Ruelas of RadiaBeam Technology. He acknowledges the photography support of E. McCrory (FNAL). He also acknowledges J. Parizek of Crytur for their executing the custom order and discussions on x-ray imaging. The first author dedicates this paper to JLL, KLL, and ELR.